# Reconfigurable THz Plasmonic Antenna Concept Using a Graphene Stack

M. Tamagnone,[1,2] J. S. Gómez-Díaz,[1] J. R. Mosig,[2] J. Perruisseau-Carrier[1, a]

[1] Adaptive MicroNano Wave Systems, LEMA / Nanolab, Ecole Polytechnique Fédérale de Lausanne (EPFL), 1015 Lausanne, Switzerland

[2] Laboratory of Electromagnetics and Acoustics (LEMA), Ecole Polytechnique Fédérale de Lausanne (EPFL), 1015 Lausanne, Switzerland

The concept and analysis of a Terahertz (THz) frequency-reconfigurable antenna using graphene are presented. The antenna exploits dipole-like plasmonic resonances that can be frequency-tuned on large range via the electric field effect in a graphene stack. In addition to efficient dynamic control, the proposed approach allows high miniaturization and good direct matching with continuous wave THz sources. A qualitative model is used to explain the excellent impedance stability under reconfiguration. These initial results are very promising for future all-graphene THz transceivers and sensors.

Graphene is currently drawing exceptional attention for electromagnetic (EM) applications at microwave, terahertz, and optical frequencies[1-6]. Since it can be considered as a true two-dimensional material, graphene is fully described by its surface conductivity $\sigma$, expressed by the Kubo formula as a function of frequency $f$, temperature $T$, chemical potential $\mu_c$ and scattering rate $\Gamma$[7, 8]. One of the most interesting properties of this material is its ability to support transverse magnetic (TM) surface plasmonic modes with unprecedented properties. Specifically, in the THz band $\sigma$ can take a highly inductive behavior, leading to very slow TM plasmonic modes that open the way to miniaturized and low-loss devices, as well as interesting sensing applications[3, 4, 6].

Moreover, graphene electrical field effect can be used to dynamically control $\sigma$ over a wide range via an applied bias voltage[3, 7, 9]. The properties of plasmonic propagation, and in particular the propagation constant, can consequently be tuned dynamically. This principle can be exploited in several ways to create devices such as modulators[9] or reconfigurable reflective elements[6]. The dynamic control of the propagation constant also enables frequency reconfiguration, as shown below.

These considerations motivate the study of the performance and reconfiguration capabilities of graphene antennas, i.e. radiators actually able to couple small THz sources, such as photomixers, to free space propagation (in contrast with graphene used as scatterers studied in fixed configuration[4]). The integration of such antennas with future graphene sources and detectors[1] will pave the way to highly efficient and integrated all-graphene THz devices.

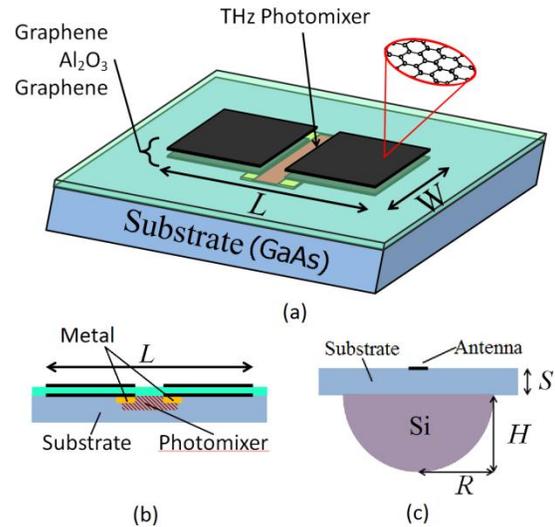

Fig. 1. Proposed graphene dipole antenna. (a) 3D view of the structure. Each half of the antenna consists of two graphene self-biasing patches separated by a thin $Al_2O_3$ layer. (b) Cross-section of the antenna. The lower graphene layer is in contact with the metallic electrodes of the source (here a THz photomixer). (c) Setup including the silicon lens for better directivity.

The graphene dipole-like antenna studied in this work is depicted in Fig. 1. Each dipole arm is a set of two stacked graphene patches separated by a thin $Al_2O_3$ insulating film ($\varepsilon_r = 9$, $\tan\delta = 0.01$). The intermediate $Al_2O_3$ layer provides a convenient way to dynamically control graphene complex conductivity via electrostatic field effect[9]. A layer thickness of 100 nm is considered, which is sufficiently large for capacitive quantum effects to be negligible[10], but thin enough to preserve low bias voltage and good coupling between the patches, as discussed next. The antenna width $W$ is 7 μm and the total length $L$ is 11 μm. The structure lies on a dielectric substrate (here GaAs, $\varepsilon_r$ = 12.9 and tan$\delta$ = 0.001) and includes a gap of 2 μm representing the THz photomixer[11].

---

[a] Author to whom correspondence should be addressed.
Email: julien.perruisseau-carrier@epfl.ch

Note that for a simpler technological implementation only the lower patches are directly connected to the photosource metallic electrodes [Fig.1(b)]. However, since both graphene layers are only separated by an electrically very thin 100 nm dielectric, they are very strongly capacitively coupled and behave as a single layer whose conductivity is approximately twice that of an individual layer, namely $\sigma_2 \approx 2\sigma_1$. This behavior is assumed in the following explanations and analytical considerations, and is verified based on full-wave simulations in section III (actually, if not stated otherwise, all full wave results shown include both graphene layers). Finally, since silicon lenses are often used with THz antennas for higher directivity, the design includes the semi-elliptical lens shown in Fig. 1(c) (S = 160 μm, H = 572 μm, R = 547 μm, $\varepsilon_r$ = 11.66, tan$\delta$ = 0.0002).

The general resonance and radiation mechanisms of the antenna are similar to those of standard metallic dipoles, except for the important difference that the standing wave on the dipole corresponds to the propagation of the plasmonic mode along graphene. Since in most cases of interest and in this work these modes are much slower than light[8], the resulting radiator is at the same time resonant and electrically very small, as further discussed later.

The propagation constant of a TM wave on an equivalent single layer of conductivity $\sigma_2 \cong 2\sigma_1$ (see above) placed on a dielectric substrate with relative permittivity $\varepsilon_{rs}$ is[8]:

$$\gamma \cong \varepsilon_0(\varepsilon_{rs}+1)\omega\sigma_2^{-1} \quad , \quad \omega = 2\pi f \qquad (1)$$

When the stack is biased, electrons are induced on one of the layer, while holes are induced on the other one. The excess of carriers alters the chemical potential $\mu_c$ of the two patches, and allows to control their conductivity and hence their plasmonic propagation constant[7]. Importantly, the symmetry of graphene band structure ensures the same upper and lower conductivities (neglecting small residual unwanted doping), and justifies the assumption $\sigma_2 \cong 2\sigma_1$ also when biasing is used. For different values of $\gamma$ the resonances of the structure geometry will occur at different frequencies, leading to an effective frequency reconfiguration technique. The $\mu_c$ range considered here is 0-0.2 eV, values that can be easily reached in practice with the proposed biasing scheme using a 0-3 V voltage range[7,9]. Finally, a typical scattering rate $\Gamma = (2\tau)^{-1}$ with $\tau$ = 1 ps is assumed in this work[7,8,12].

The proposed antenna was simulated using Ansys HFSS. The corresponding input impedance over the considered 0-0.2 eV range of chemical potential is shown in Fig. 2.

Different interesting features can be observed. First, the real part of the input impedance is large at the resonance frequency, namely about 500 Ω. This is a very desirable feature for minimizing return loss when connecting the antenna to a photomixer (since such devices have high impedance), a need emphasized in literature regarding different THz applications[13,14]. Second, the resonant frequency $f_r$ can be controlled over a very wide range from 0.8 THz to 1.8 THz. Remarkably, the impedance itself remains very stable upon reconfiguration, thereby avoiding the need for a lossy and complex reconfigurable matching network. Finally, it is worth mentioning that the smaller secondary peaks in Fig. 2 are due to the first higher resonance of the structure in its longitudinal dimension. The fact that the frequency of the secondary peaks is not twice that of the main resonances is due to the dispersive nature of the plasmonic propagation discussed as shown in (4) and illustrated in Fig. 4. Moreover, it was shown that the location of both primary and secondary resonances are almost unaffected by the antenna transversal dimension W.

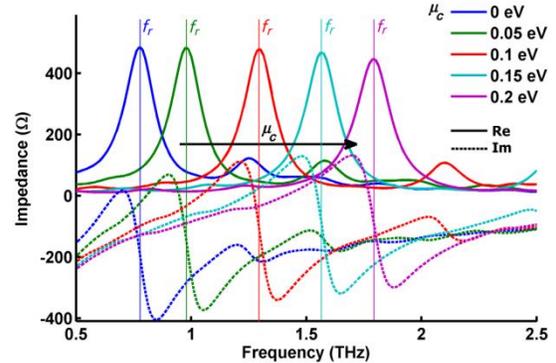

Fig. 2. Real and imaginary parts of the antenna input impedance upon reconfiguration.

In the remainder of this section we provide a qualitative yet insightful analysis of this behavior, based on the transmission line (TL) model of Fig. 3 and approximate analytical expressions related to the plasmonic propagation on graphene.

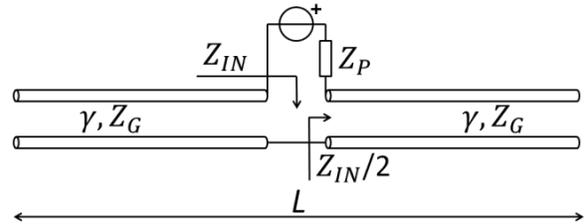

Fig. 3. Qualitative transmission line model of the antenna.

In the considered frequency range the Kubo formula for graphene conductivity can be written in the form[7]:

$$\sigma_1^{-1} = (2\Gamma + j\omega)g(\mu_c, T) \quad (2)$$

where $g(\mu_c, T)$ is a real function independent of frequency given by:

$$g(\mu_c, T) = \frac{\pi \hbar^2}{q_e^2 k_B T}\left[\frac{\mu_C}{k_B T} + 2\ln\left(1 + e^{-\frac{\mu_C}{k_B T}}\right)\right]^{-1} \quad (3)$$

Combining (1) and (2) and using the $\sigma_2$ approximation explained in section II B:

$$\gamma = \alpha + j\beta \cong \frac{1}{2}\left(\frac{2\Gamma}{\omega} + j\right)\varepsilon_0(\varepsilon_{rs}+1)\omega^2 g(\mu_c,T) \quad (4)$$

Equation (4) shows that both the real and the imaginary part of $\gamma$ depend on frequency and chemical potential $\mu_c$, and that their ratio depends on frequency as follows:

$$\alpha/\beta = 2\Gamma/\omega \quad (5)$$

Now, let us define $f_r(\mu_c)$ as the antenna resonant frequency, i.e. the frequency where the real part of the antenna input impedance $Z_{IN}$ is maximum (see Fig. 2). Similarly, the subscript '$r$' is used to explicit the value of $\alpha$, $\beta$, and the antenna impedance $Z_{IN}$ at the resonant frequency $f_r$. The antenna was designed for resonance at $\beta = \beta_r \equiv 2\pi/L$, and therefore the impedance $Z_{IN,r}$ writes[15]:

$$Z_{IN,r} = 2\frac{Z_G}{\alpha_r L/2} = \frac{4Z_G}{\alpha_r L} \quad (6)$$

where $Z_G$ is the characteristic impedance of the plasmonic mode corresponding to the TL sections in the model. The link between $f_r$, $\beta(f, \mu_c)$, $\beta_r$, and $\mu_c$ is illustrated in Fig. 4 for intuitive understanding.

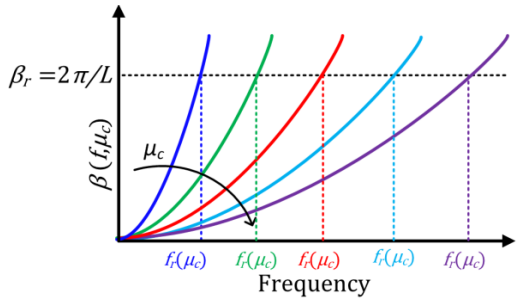

Fig. 4. Illustration of the resonant frequency dynamic tuning.

Since the supported mode is TM, $Z_G$ can be expressed as[15] $Z_G = \beta/(\omega\varepsilon_{eff})$. This expression can be introduced in (4), yielding:

$$Z_{IN,r} = \frac{4\beta_r/(\omega_r\varepsilon_{eff})}{\alpha_r L} = \frac{4\beta_r/\alpha_r}{L\omega_r\varepsilon_{eff}} = \frac{2}{\Gamma L\varepsilon_{eff}} \quad (7)$$

Equation (7) shows that $Z_{IN,r}$ is independent of frequency. In other words, this result corresponds to the full-wave based observation made in Fig. 2 that the impedance of the antenna at its resonance frequency remains constant upon reconfiguration. Intuitively, this is due to the particular frequency dependence between α and β expressed in (5).

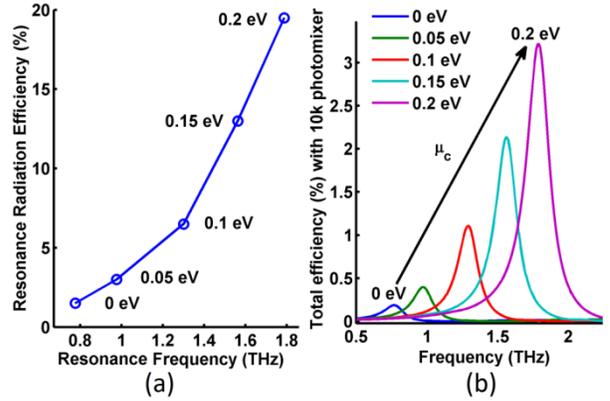

Fig. 5. (a) Radiation efficiency at the resonance versus resonant frequency for different values of $\mu_c$. (b) Total efficiency (including matching and radiation) of the antenna assuming a 10 kΩ photomixer.

Fig. 5(a) shows the achieved radiation efficiency $\eta$ at the resonant frequency $f_r(\mu_c)$ when the chemical potential $\mu_c$ is varied. The rather low values of $\eta$ are linked with the very small electrical size of the antenna, namely $L \approx \lambda_0/34$ at 0.8 THz and $L \approx \lambda_0/15$ at 1.8 THz. It is also observed that the radiation efficiency improves for larger $\mu_c$, as expected since the antenna becomes electrically larger at resonance

Fig. 5(b) plots the antenna total efficiency when fed by a THz photomixer of internal impedance $Z_P = 10$ kΩ (which is a typical value for continuous wave systems[11]). The performance is comparable with common metal implementations, where the impedance mismatch alone often limits the total efficiency to a few percent[14]. Resonant metal implementations[13] can achieve total efficiencies in the order of 20 % in a narrow band, but they do not allow reconfigurability and are approximately 20 times larger than the graphene-based antenna proposed here.

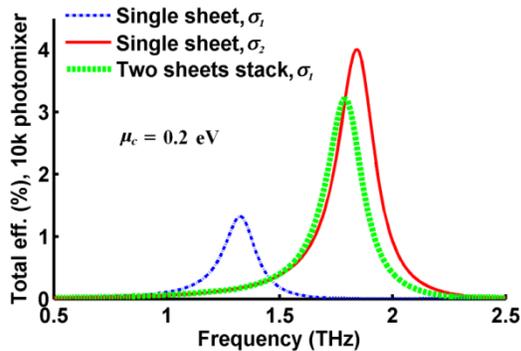

Fig. 6. Total efficiency obtained by simulating the actual two-layer graphene structure of Fig. 1, compared with a single layer of equal and double conductivity, respectively.

Fig. 6 compares the total efficiency (a good representation of both matching and loss issues) of the actual double layer antenna with that obtained by replacing the two layers by a single layer of the same and double conductivity ($\sigma_1$ and $\sigma_2$ respectively). It is observed that the single-layer simulation with $\sigma_2$ is very similar to the complete two layer simulation with $\sigma_1$, which justifies the assumption used in Section II B for the qualitative model. By contrast, as expected, a single sheet of conductivity $\sigma_1$ exhibits shifted resonance frequency and different maximum efficiency.

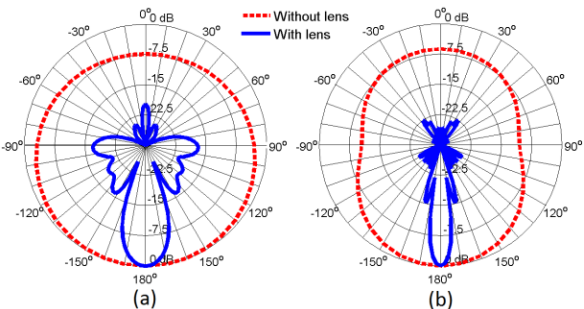

Fig. 7. Directivity (H-plane) at the resonance with and without lens. (a) $\mu_c = 0$ eV, f = 0.8 THz. (b) $\mu_c = 0.2$ eV, f = 1.8 THz.

Fig. 7 shows the normalized directivity of the antenna with and without the silicon lens shown in Fig. 1(c), confirming the expected increase in directivity. It is also worth mentioning that, as already observed in similar cases[13], the lens has negligible impact on the antenna input impedance.

It has been shown that graphene exhibits extremely interesting properties for the design of THz antennas, much beyond their use as simple scatterers. In particular, our results demonstrate that simple structures can already provide excellent properties in terms of high input impedance, frequency-reconfiguration, and stable impedance and radiation patterns upon tuning. These initial results are extremely promising for the future monolithic integration of antenna and THz emitter/detector based on graphene.

## REFERENCES

1. Z. Zixu, S. Grover, K. Krueger and G. Moddel, presented at the Photovoltaic Specialists Conference (PVSC), 2011 37th IEEE, 2011.
2. H. Yi, W. Lin-Sheng, T. Min and M. Junfa, Nanotechnology, IEEE Transactions on **11** (4), 836-842 (2012).
3. L. Vicarelli, M. S. Vitiello, D. Coquillat, A. Lombardo, A. C. Ferrari, W. Knap, M. Polini, V. Pellegrini and A. Tredicucci, Nature Materials **11**, 865–871 (2012).
4. I. Llatser, C. Kremers, A. Cabellos-Aparicio, J. M. Jornet, E. Alarcón, D. N. Chigrin, Photon Nanostruct: Fundam Appl **10** (4), 353-358 (2012)
5. J. S. Gomez-Diaz, J. Perruisseau-Carrier, P. Sharma and A. Ionescu, Journal of Applied Physics **111** (11), 114908-114907 (2012).
6. J. Perruisseau-Carrier, in *2012 Loughborough Antennas & Propagation Conference* (Loughborough, UK, 2012).
7. G. W. Hanson, Antennas and Propagation, IEEE Transactions on **56** (3), 747-757 (2008).
8. M. Jablan, H. Buljan and M. Soljačić, Physical Review B **80** (24), 245435 (2009).
9. M. Liu, X. Yin and X. Zhang, Nano Letters **12** (3), 1482-1485 (2012).
10. C. Zhihong and J. Appenzeller, presented at the Electron Devices Meeting, 2008. IEDM 2008. IEEE International, (2008).
11. I. S. Gregory, C. Baker, W. R. Tribe, I. V. Bradley, M. J. Evans, E. H. Linfield, A. G. Davies and M. Missous, Quantum Electronics, IEEE Journal of **41** (5), 717-728 (2005).
12. P. Neugebauer, M. Orlita, C. Faugeras, A. L. Barra and M. Potemski, Physical Review Letters **103** (13), 136403 (2009).
13. I. Woo, T. K. Nguyen, H. Han, H. Lim and I. Park, Opt. Express **18** (18), 18532-18542 (2010).
14. H. Yi, N. Khiabani, S. Yaochun and L. Di, presented at the Antenna Technology (iWAT), 2011 International Workshop on, (2011).
15. R. E. Collins, *Foundations for Microwave Engineering*. (McGraw-Hill, 1966).